\documentclass[10pt]{iopart}
\usepackage{iopams} 
\expandafter\let\csname equation*\endcsname\relax
\expandafter\let\csname endequation*\endcsname\relax\usepackage{subfigure}
\usepackage{amsmath}
\usepackage{epsfig}
\usepackage{graphicx}
\usepackage[usenames,dvipsnames]{color}
\usepackage[utf8]{inputenc}
\usepackage{tikz}
\usepackage{cite}
\usepackage{lineno}
\usepackage[hidelinks,unicode=true]{hyperref}
\hypersetup{colorlinks=true,
	linkcolor=blue,
	urlcolor=blue,
	citecolor=blue,
	pdfhighlight=/N
}

\usepackage{comment}
\usepackage[normalem]{ulem}

\begin{document}

\title{Radial Evolution in a Reaction-Diffusion Model}
\author{Sofia M. Silveira and Sidiney G. Alves}
\address{Departamento de Estat\'istica, F\'isica e Matem\'atica, Universidade Federal de S\~ao Jo\~ao Del-Rei,
36420-000 Ouro Branco, Minas Gerais, Brazil}
\begin{abstract}
	In this work, we investigate an off-lattice version of the diffusion-reaction model, $A + A \leftrightarrow A$.  We consider extensive numerical simulation of the radial system obtained from a single seed. Observed fluctuations in such an evolving system are characterized by a circular region occupied by particles growing over an empty one. We show that the fluctuating front separating the two regions belongs to the circular subclass of the Kardar-Parisi-Zhang universality class.
\end{abstract}



\section{Introduction}
Reaction-diffusion (RD) processes are widespread across various fields, ranging from chemistry to biology. In some of these systems, the interplay between reaction and diffusion mechanisms leads to a traveling wave invasion of an occupied or stable region into an empty or unstable one \cite{Riordan1995}. The evolution of the system gives rise to a highly distinct interface that effectively separates the two regions, thereby defining a front propagation that evolves far from equilibrium.

The fluctuations observed in such an evolving interface are characterized by self-similarity and universality that emerge from distinct dynamical processes of their formation \cite{barabasi,meakin}, a subject of significant interest in the statistical physics field. A well-studied model that exhibits front propagation between an occupied and an empty region is the $A + A \leftrightarrow A$ RD model \cite{BenAvraham1990}. 
Most studies conducted with this model have been focused on the scaling exponents of the front width or roughness (the standard deviation) \cite{Riordan1995,Tripathy2000,Tripathy2001,Moro2001,Moro2004,Nesic2014}.
The obtained exponent in the most recent works \cite{Tripathy2000,Tripathy2001,Moro2001,Moro2004,Nesic2014} indicates that the evolution belongs to the Kardar-Parisi-Zhang (KPZ) universality class. The KPZ universality class is one of the most studied classes. It is associated with the equation proposed by Kardar, Parisi, and Zhang \cite{KPZ}
\begin{eqnarray}
	\frac{\partial h(x,t)}{\partial t} = \nu \nabla^2 h(x,t) + \lambda \left( \nabla h(x,t) \right)^2 + \eta(x,t)
\end{eqnarray}
where $h(x,t)$ represents the interface height at a position $x$ in time $t$, the first term on the right-hand side is related to surface tension, the non-linear second term represents a local lateral growth in the normal direction along the surface and the last one is a white noise with $\langle \eta(x,t)\rangle = 0$ and  $\langle \eta(x,t) \eta(x',t')\rangle = D \delta(x-x')\delta(t-t')$.
The KPZ universality class is associated with a wide range of systems that exhibit non-equilibrium fluctuations, spanning from film deposition \cite{Almeida2014,Almeida2015} to biological growth \cite{Huergo2010,Huergo2011,Huergo2012}. Significant progress has been made in understanding the KPZ universality class over the past decades \cite{johansson,PraSpo1}.
Considering the non-stationary regime, the height fluctuation of a single point is governed by \cite{krug92}
\begin{equation}
	h = v_\infty t+s_\lambda(\Gamma t)^{\beta}\chi,
	\label{eq:hdet}
\end{equation}
where $v_\infty$ is the asymptotic growth velocity, $s_\lambda$ is the sign of the parameter $\lambda$ in the KPZ equation, $\Gamma$ is a parameter related to the amplitude of height fluctuation, $\beta$ is the growth exponent and $\chi$ is a stochastic variable.
It was demonstrated that the probability distribution function (PDF) of $\chi$ depends on the initial condition geometry and boundary condition \cite{PraSpo1,PraSpo2,Corwin2012,Takeuchi2018}. For the one-dimensional case, the PDFs are given by the Tracy-Widom  PDF of the Gaussian orthogonal ensemble (GOE) in the case of flat or fixed size substrates or by the Tracy-Widom PDF from a Gaussian unitary ensemble (GUE) for circular one or with the enlarging size \cite{PraSpo1,PraSpo2}. Therefore, these results split the KPZ universality class into two, namely flat and circular KPZ subclasses.
A large number of theoretical \cite{SasaSpohn,Amir,Calabrese,Imamura}, experimental \cite{TakeSano,TakeuchiSP,TakeSano2012}, and numerical  \cite{Alves11,Oliveira12,Alves13,Carrasco_2014,Healy14,Santalla_2015,HealyTake,Santalla_2017,Alves2018,Roy} works confirmed the universality of the fluctuation's PDF and its dependency on the geometry.

In a recent work, Barreales et al. \cite{Barreales_2020} conducted a numerical investigation of the spatio-temporal fluctuations of the front propagation in the $A + A \leftrightarrow A$ reaction-diffusion model with a planar initial condition.
They focused on analyzing the kinetic roughness associated with the one-dimensional front propagation, employing the well-established framework of the KPZ universality class.
Their findings corroborate that the fluctuations observed in the front propagation belong to the KPZ universality class. 
Specifically, they observed behaviors consistent with the Tracy-Widom PDF from the GOE of the flat KPZ subclass.
In the present work, we focus on an off-lattice variant of the same reaction-diffusion model, where the initial condition is given by a single seed at the system's origin. As the model evolves, it gives rise to a dynamic wherein particles are confined within a circular region centered at the system's origin. The density of particles and the size of fluctuations in the region's front depend on the control parameter. Through careful simulations, we demonstrate that the front propagation fluctuation conforms to the KPZ universality class. We obtained a behavior that agrees with the Tracy-Widom PDF from the GUE of the circular KPZ subclass.

The rest of this article is organized as follows. The next Section provides a detailed description of the model and simulational strategies employed. The findings and discussions are presented in Section \ref{results}. Finally, in Section \ref{conclusions} the conclusions are drawn.

\section{Model and Methods}

We explore the dynamics of front particles in an off-lattice version of the radial symmetric RD model. In our investigation, we consider particles with a diameter of size $a=1$ in a two-dimensional uniform region. The simulations begin with a single particle located at the system's origin.

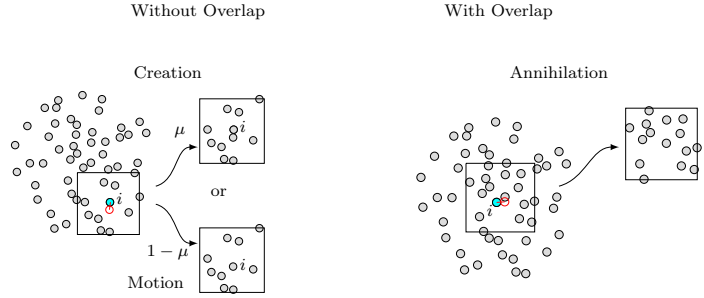
\begin{figure}[t]
	\begin{indented}
		\item[]	
		\scalebox{0.8}{\begin{tikzpicture}
				\node at (0,3) {Without Overlap};
				\node at (5,3) {With Overlap};
				\node at (-1,0) {\includegraphics*[width=0.4\linewidth]{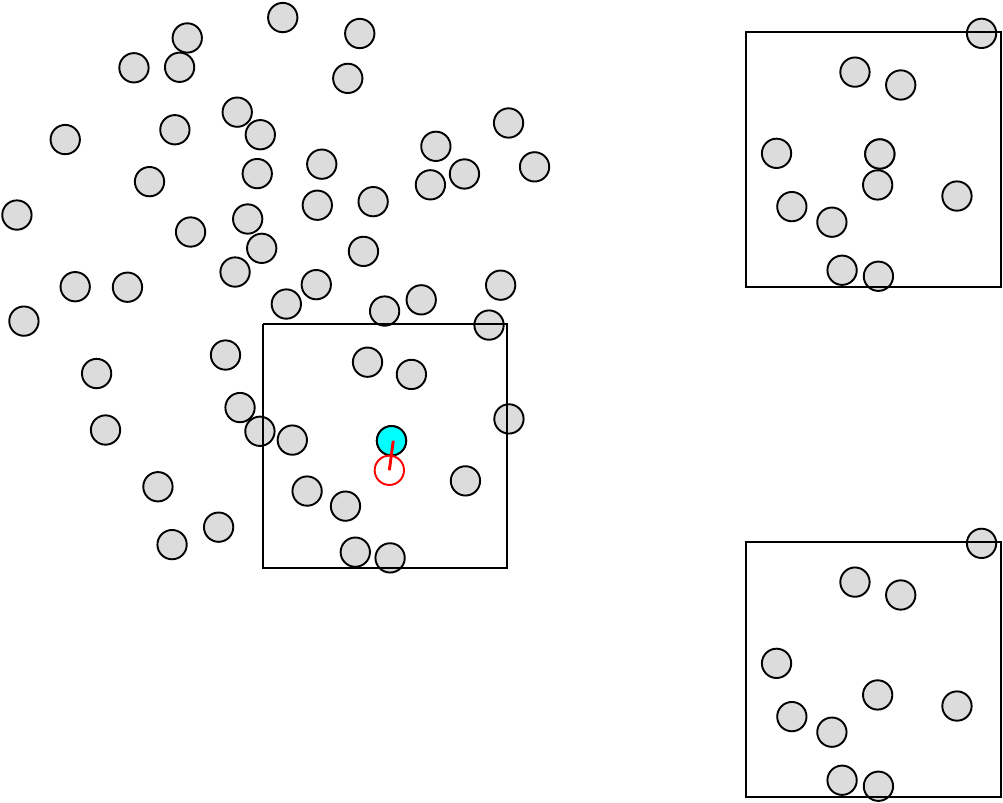}};
				\node at (6,0) {\includegraphics*[width=0.45\linewidth]{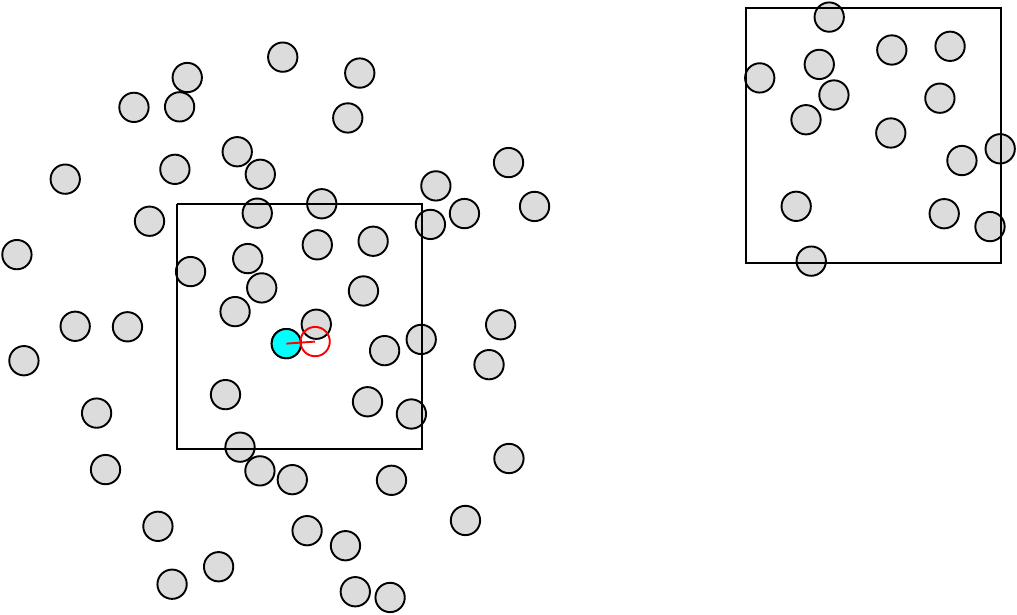}};
				\node at (0.35,0) {or}; 
				\draw[-latex] (-0.7,0.1) to [out=15,in=180] (0,0.75); \draw[-latex] (-0.7,-0.2) to [out=-15,in=180] (0,-0.85);
				\node at (-0.3,1) {$\mu$}; \node at (-0.5,-1) {$1-\mu$};
				\node at (-0.5,2) {Creation}; \node at (-0.7,-1.5) {Motion};
				
				\draw[-latex] (6,0.1) to [out=15,in=180] (7,0.75); 
				
				
				\node at (6,2) {Annihilation}; 
				\node at (-1.3,-0.1) {$i$};\node at (0.75,1.1) {$i$};\node at (0.75,-1.2) {$i$};
				\node at (4.85,-0.3) {$i$};
		\end{tikzpicture}}
	\end{indented}
	
	\caption{\label{fig:growth_rule}The growth rule can be illustrated in two different scenarios. In the first scenario (left panel), a random direction is chosen (represented by the open red circle), and it is checked whether there is any particle adjacent to the selected particle $i$ (highlighted in cyan). If there are no adjacent particles, a new particle is added to the random direction with a probability $\mu$ (top highlighted square). Alternatively, with the complementary probability, the particle $i$ is moved to occupy that position (bottom highlighted square). In the second scenario (right panel), when there is already a particle in the random direction, the particle $i$ is deleted.}
\end{figure}

The evolution rules are implemented as follows. At each time step, a particle $i$ and a direction $\phi$ are randomly selected. It is then checked if placing a new particle adjacent to the particle $i$ (in the $\phi$ direction) would overlap with another particle, as illustrated in Figure \ref{fig:growth_rule}. If no overlap occurs, two actions are possible (left panel in Figure \ref{fig:growth_rule}). In the first action, with a probability $\mu$, a new particle is created adjacent to the particle $i$ at the $\phi$ direction.
The second action involves the motion of the particle $i$ to this position with the complementary probability. When an overlap occurs (right panel in Figure \ref{fig:growth_rule}), the selected particle is removed. The time is incremented by $\Delta t=1/N$ at each attempt, where $N$ is the total number of particles in the system.

The simulations were conducted on regions that permitted the system to evolve, reaching a radius of up to $4000a$. Up to $10^4$ independent realizations were used. To enhance the efficiency and computational performance of the simulations, enabling faster data generation and analysis,
we employed optimization strategies as described in \cite{AlvesBJP}\footnote{The computer time demanded, on a computer equipped with a processor Intel(R) Xeon(R) CPU X5690@3.47GHz processor, to simulate of just one sample, considering $\mu =0.1$ (to the system reach a radius of 2500a) and $0.90$ (reach a radius of 3000a) were approximately 4 and 5 hours, respectively.}. 

\section{Results and discussions}
\label{results}
\begin{figure}[t]
\begin{indented}
	\item[]	\includegraphics*[width=0.65\linewidth]{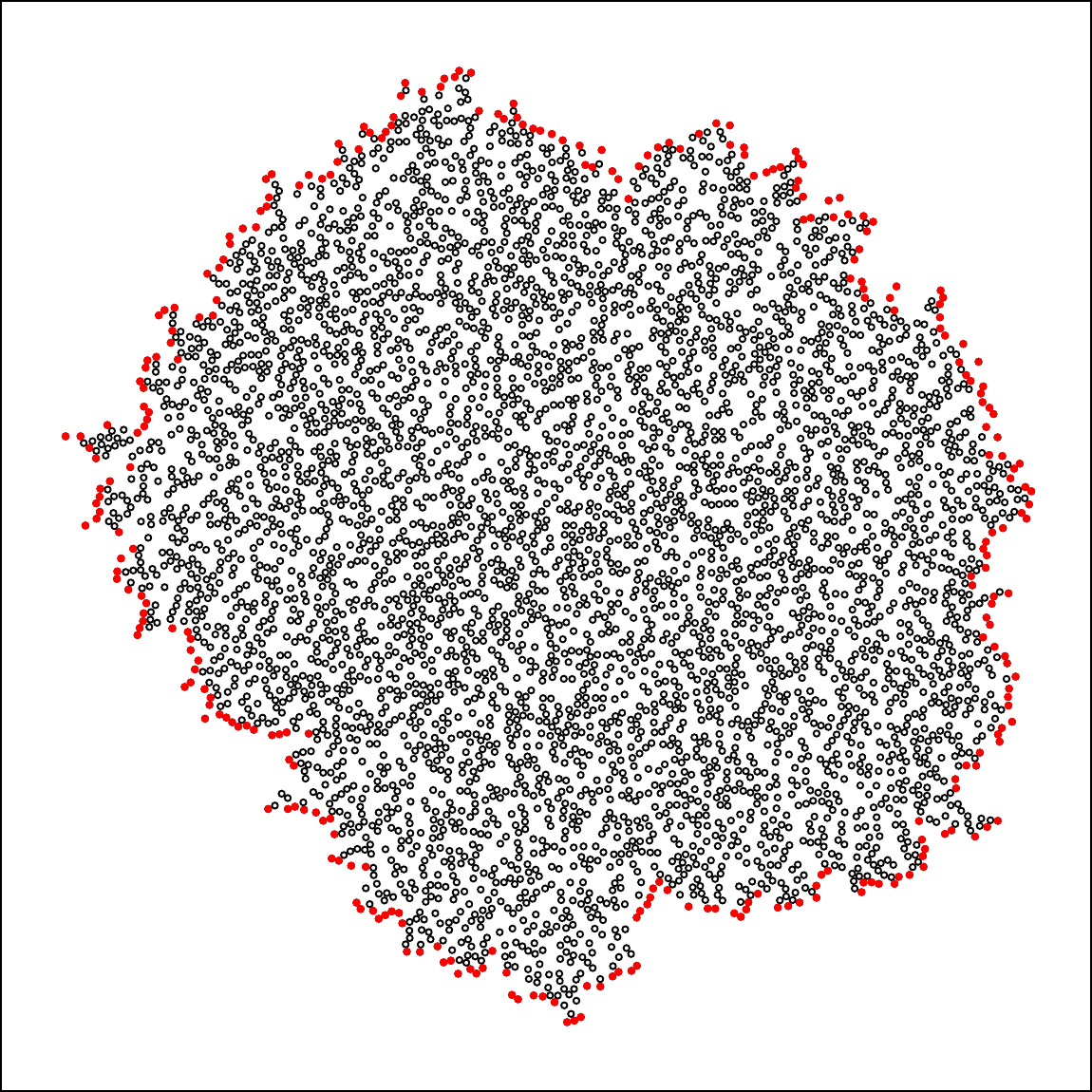}
\end{indented}
	\caption{\label{fig:snapshot} Snapshot obtained using $\mu=0.2$. The radius of the occupied region is $200a$ with about 5000 particles (the bounding box size is $400a$). The most external particles are depicted in red.}
\end{figure}

~
The system evolution leads to a set of particles distributed around its origin, as shown in Figure \ref{fig:snapshot}. 
This occupied region invades the empty one, establishing a well-defined fluctuating front between them. To define the particles belonging to this front, we adopt the following strategy. The circular region ($\in(0,2\pi)$) centered at the origin is discretized into $n_d = \mbox{INT}(2\pi R_a)$ intervals of angular length $\Delta \phi =\frac{1}{R_a}$, where $R_a$ is the position of the most distant particle to the system origin divided by $a$ (the particle diameter). The particle with the greatest distance from the system origin within each interval $(\phi, \phi+\Delta \phi)$ is selected to define the front. Figure \ref{fig:snapshot} depicts particles of the front (red circles), obtained using this strategy.

\begin{figure}[h]
\begin{indented}
	\item[]	
	\includegraphics[width=0.326\linewidth]{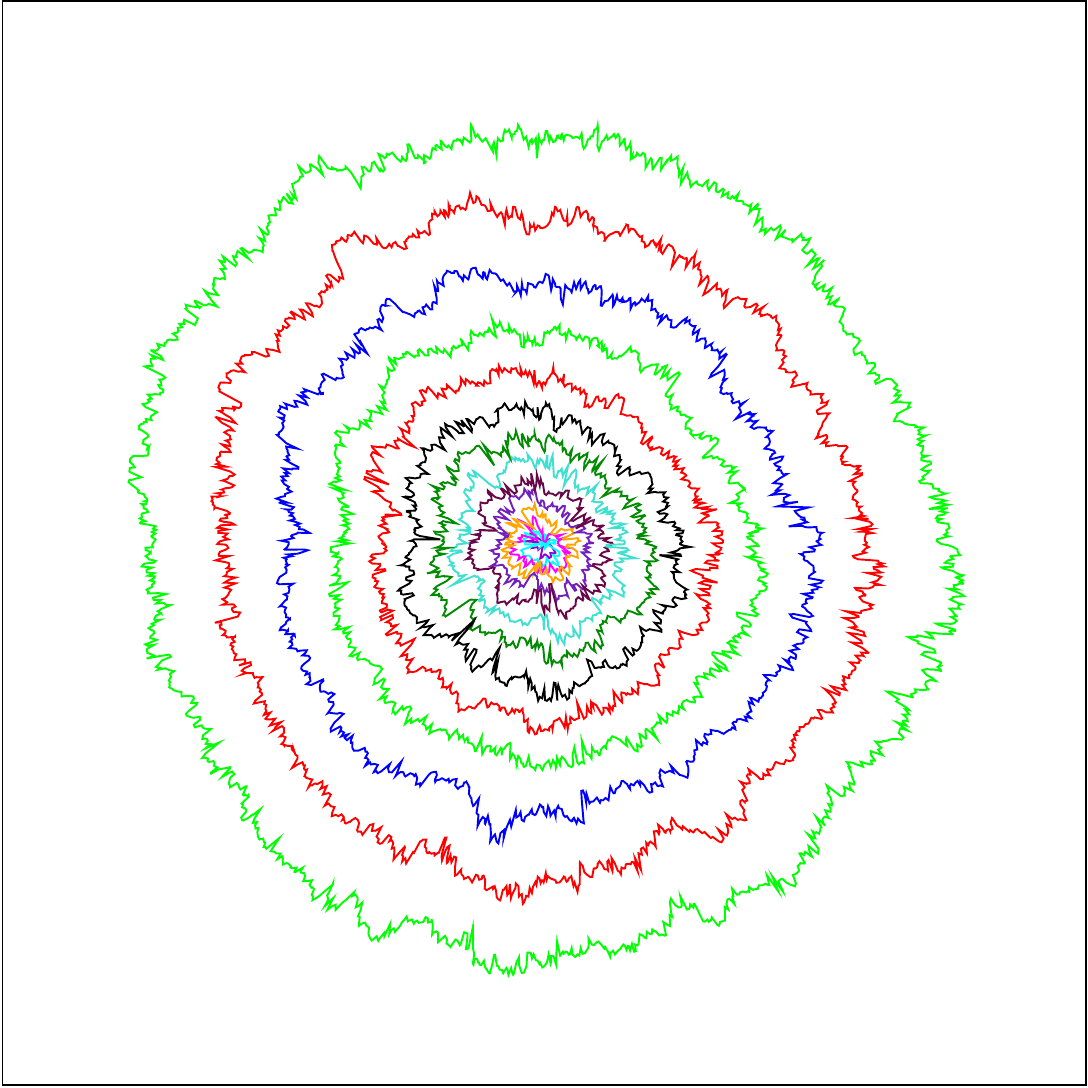}
	\includegraphics[width=0.326\linewidth]{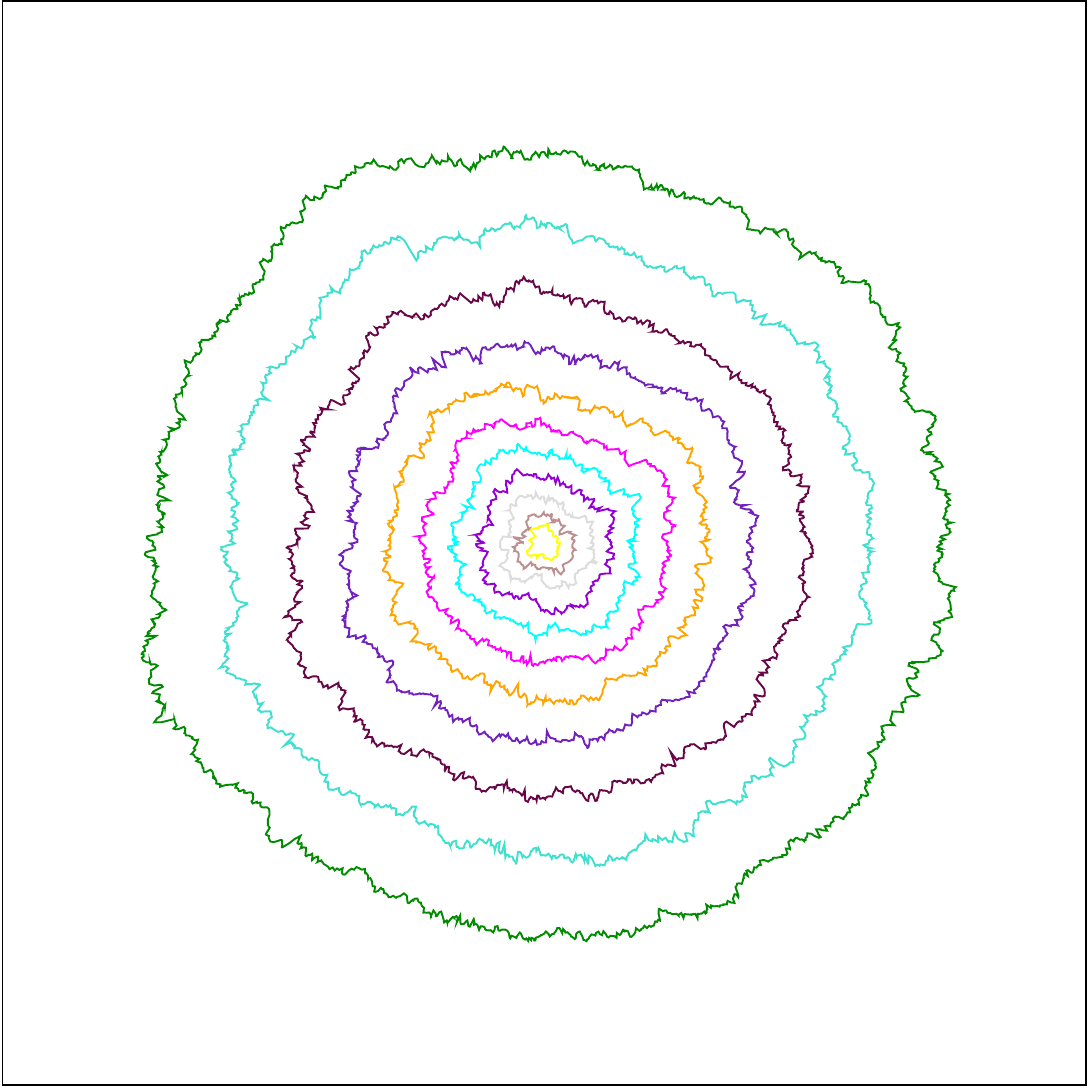}
	\includegraphics[width=0.326\linewidth]{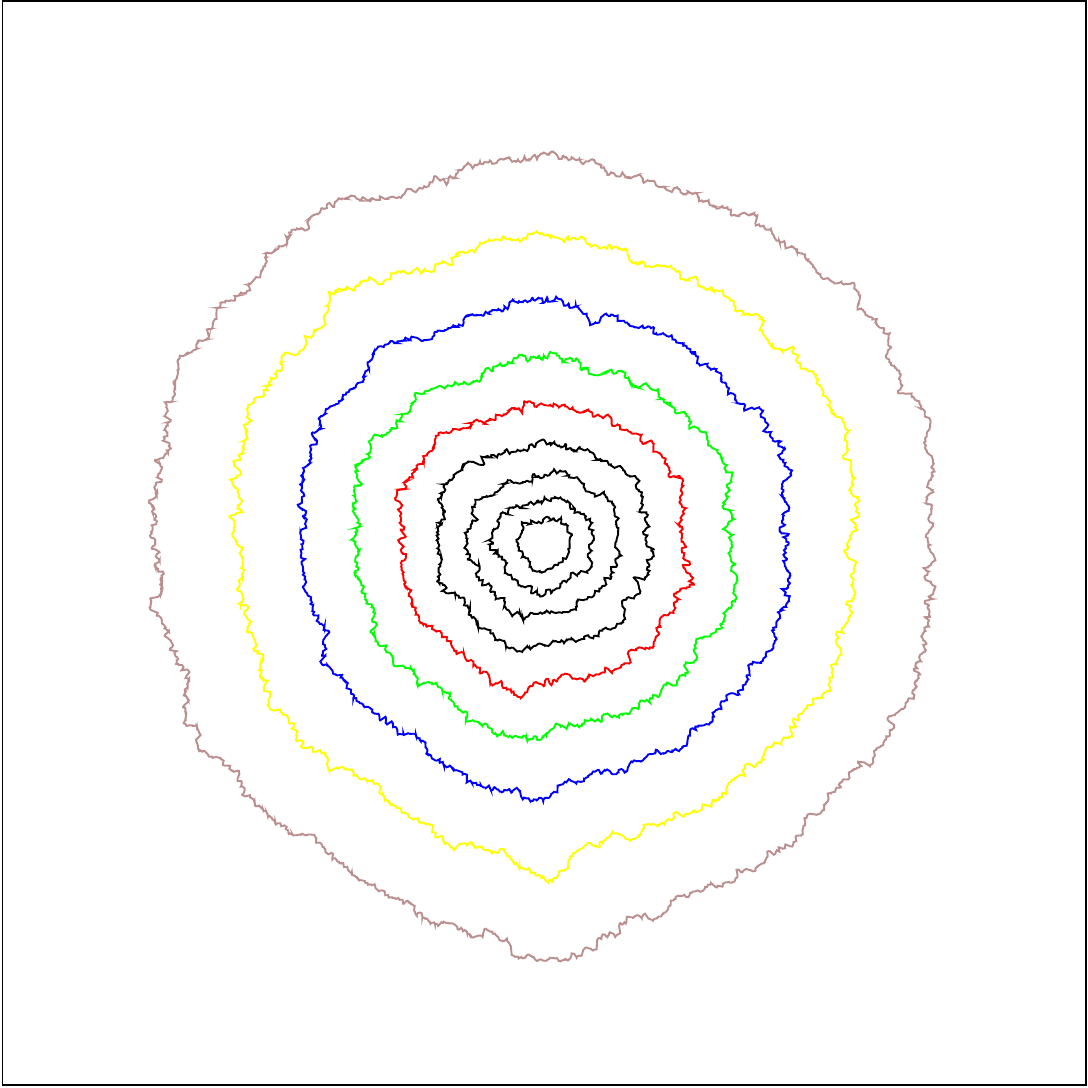}
\end{indented}
	\caption{\label{fig:boundaries}A temporal evolution snapshot of the boundaries was obtained using the probability $\mu=0.1$ (on the left), 0.3 (center), and 0.9 (right). The squares in all panels are of size equal to 2000.}
\end{figure}

The evolution of the front is depicted at different stages in Figure \ref{fig:boundaries} for three different values of the parameter $\mu$. It is worth noting that larger fluctuations are observed for smaller values of $\mu$. In this case, the regime
is mainly characterized by diffusion processes.
This illustration effectively highlights the role played by the parameter $\mu$ in shaping the separating front dynamic.

\begin{figure}[h]
	\begin{indented}
		\item[]	\includegraphics[width=0.65\linewidth]{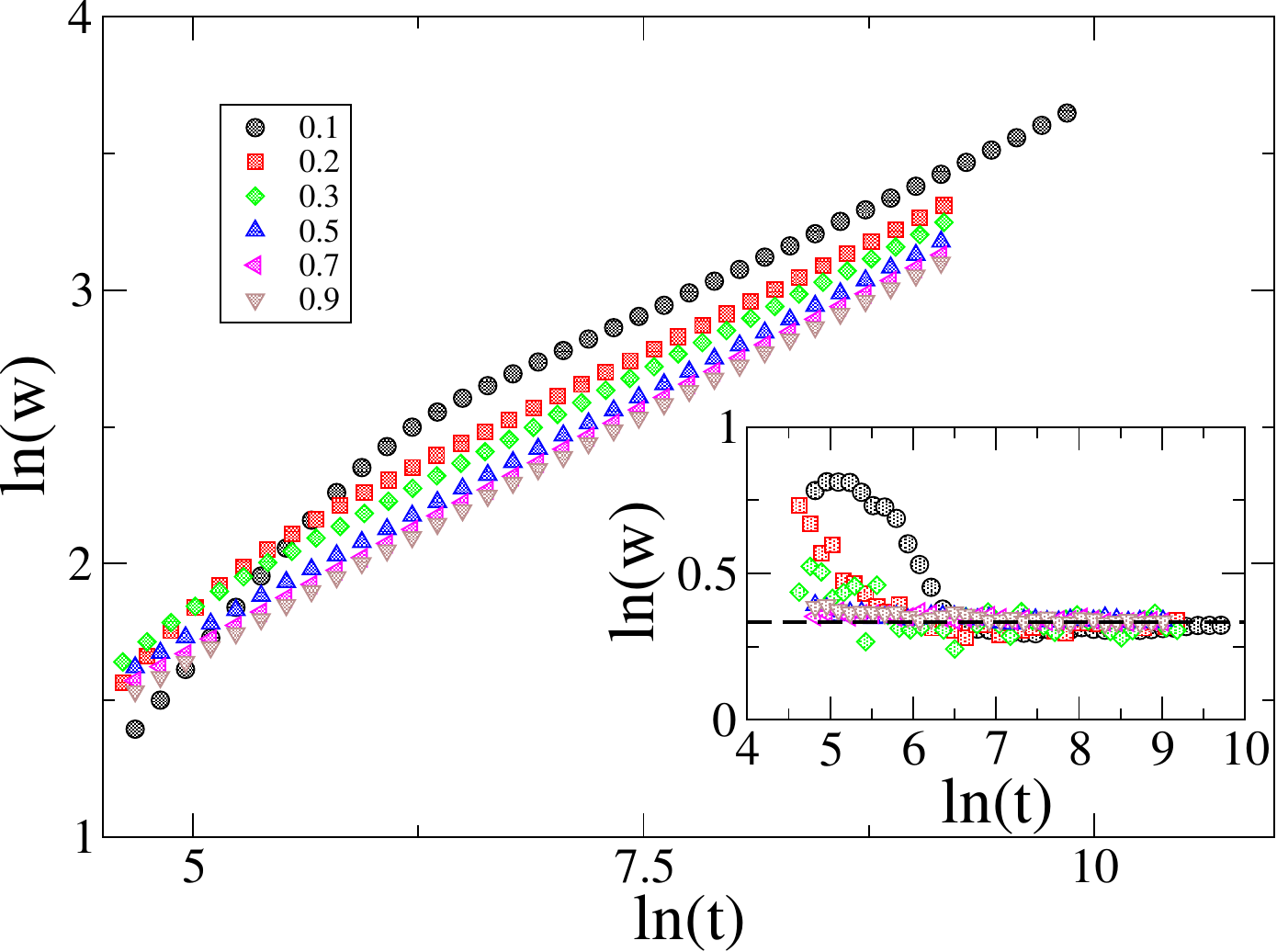}
	\end{indented}
	\caption{\label{fig:w}Roughness time evolution (main panel) and growth exponent (inset panel) for different values of the parameter $\mu$ (see the legend).  The dashed line represents the growth exponent $\beta=1/3$ expected for the KPZ universality class.} 
\end{figure}

To characterize the temporal evolution and the fluctuations of the front, we analyze the set of distances from the system's origin of the particles belonging to the front represented by $r_i(t)$ with $i=1, 2, \ldots, n_d$, here $n_d$ is the number of particles in the front (number of different angular directions used to define the boundary of the system).
We expect that a point of the front obeys the ansatz in Equation \ref{eq:hdet}. So, the distance to the origin of a particle $i$ of the front can be expressed as 
\begin{equation}
	r_i(t) = v_\infty t + s_\lambda (\Gamma t)^\beta \chi. \label{eq:point}
\end{equation}
In the next steps, we will determine the parameters $v_\infty$, $\Gamma$, and $\beta$, as well as the properties of the stochastic variable $\chi$.
We first address the determination of the growth exponent $\beta$ using the time evolution of the second cumulant or roughness. The roughness can be defined as follows
\begin{equation}
	\label{eq:w}
	w^2(t)=\left<{r^2}\right>_c=\left< \dfrac{1}{n_d}\sum_i^{n_d} \left(r_i(t)-\overline{r}(t)\right)^2\right>
\end{equation}
where  $\overline{r}(t)$ is the mean distance of the set $\{r_i(t)\}$ of the particle belonging to the front at time $t$ and $\left< \right>$ represents a mean over various samples (runs). 
Figure \ref{fig:w} shows the results obtained for different $\mu$ parameter values.
After an initial transient, the roughness exhibits a scaling following a power law behavior with approximately the same growth exponent. The inset of Figure \ref{fig:w} shows the growth exponent as a function of time. 
The $\beta$ values in Table \ref{tab:results} were obtained considering a mean in the last decade of the curves.
Note that the case with $\mu=0.1$ exhibits the largest deviation (less than $4\%$) from the expected growth exponent of the one-dimensional KPZ universality class. 

\begin{figure}[t]
\begin{indented}
	\item[]	\includegraphics[width=0.65\linewidth]{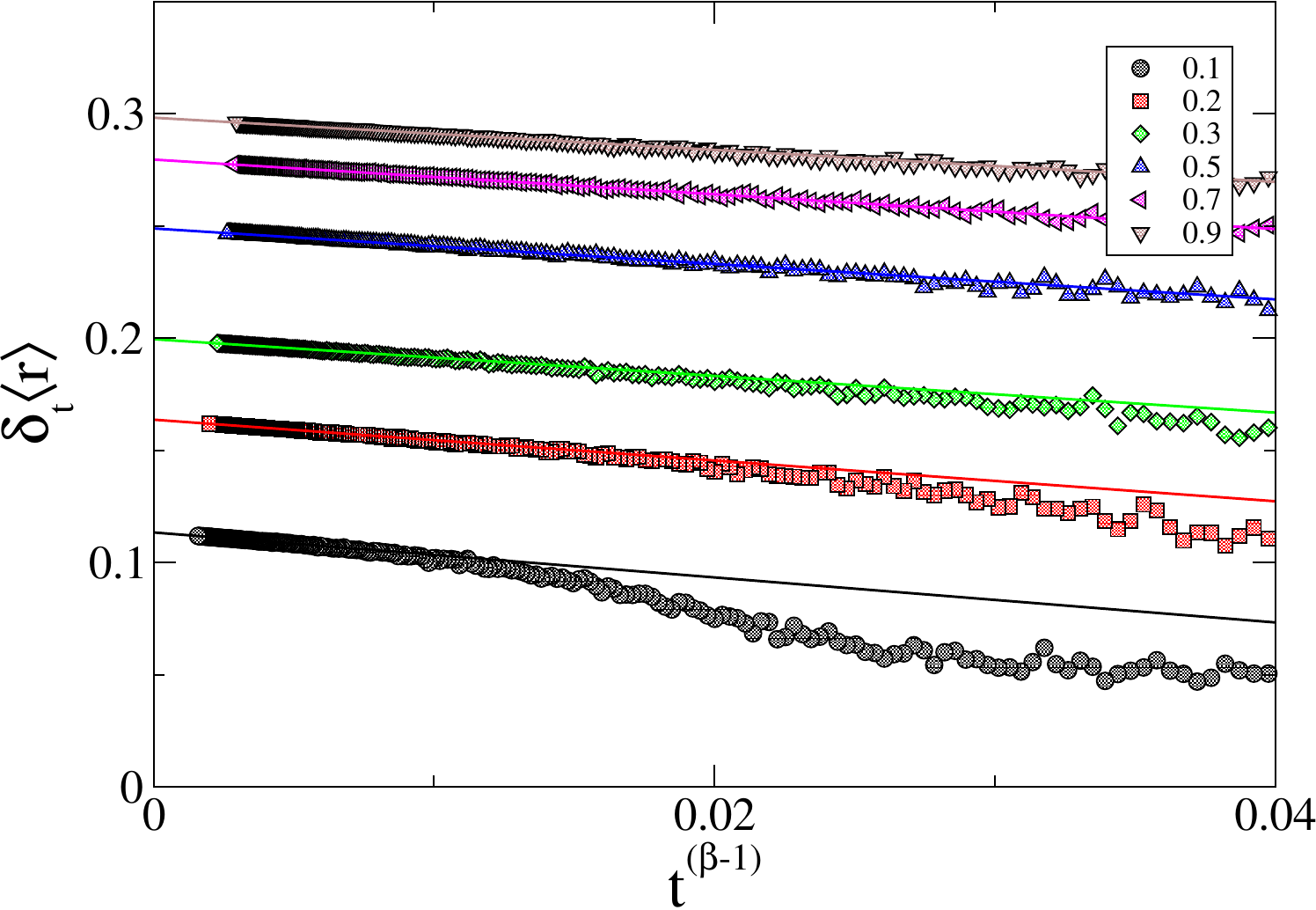}
\end{indented}
	\caption{\label{fig:vinf}Growth velocity as a function of the $t^{\beta-1}$ for distinct value of the parameter $\mu$. We extrapolate to $t\rightarrow\infty$ limit to estimate the asymptotic growth velocity.} 
\end{figure}

Now we consider determining the growth velocity using the time derivative of Equation \ref{eq:point}, {\it i.e.},
\begin{equation}
	\frac{\mathrm{d} \left<r\right>}{\mathrm{d}t} = v_\infty + s_\lambda \beta \Gamma^\beta t^{\beta - 1} \left<\chi\right>,
\end{equation}
the asymptotic value is obtained considering an extrapolation in the plot of $\frac{\mathrm{d} \left<r\right>}{\mathrm{d}t}$ against $t^{\beta - 1}$ with  $\beta=1/3$, the expected growth exponent in the one-dimensional KPZ class. The Figure \ref{fig:vinf} presents the results for different values of $\mu$. After an initial transient, we observe a linear behavior expected for systems following the ansatz in Equation \ref{eq:hdet}.
It is worth noting that the initial transient is larger for smaller values of the $\mu$ parameter. Besides, an increase in the value of $\mu$ leads to an increase in the growth velocity. As $\mu$ becomes larger, the system experiences higher particle creation probability. This increased likelihood of particle creation accelerates the growth of the occupied region, resulting in a higher growth velocity overall. This behavior agrees with the reported results for the on-lattice case with flat initial geometry \cite{Moro2001,Barreales_2020}.
The obtained values of the growth velocity are presented in Table \ref{tab:results}.

\begin{table}
\begin{indented}
	\item[]	\begin{tabular}{c||c|c|c|c|c|c|c}
		$\mu$    &  $v_\infty$ & $\beta$ & $\Gamma$ &  $\langle\eta \rangle$&  $S$ & $K$  & $A$     \\ \hline\hline
		0.1     & 0.1134(6) &  0.32(2)  &  5.3(2) &  1.12(5) &   0.20(3)    &  0.06(3)    &  70(2)  \\ \hline
		0.2     & 0.1636(4) &  0.33(2)  &  3.7(2) &  1.87(4) &    0.21(2)    &  0.08(2)   &  28(1)   \\ \hline
		0.3     & 0.1995(3) &  0.328(5)   &  2.9(3) &  1.45(5) &    0.20(2)    &  0.08(2) &  19.2(5) \\ \hline
		0.5     & 0.2488(3) &  0.336(4)   &  2.3(2) &  5.52(4) &    0.21(1)    &  0.09(1) &  12.1(3) \\ \hline
		0.7     & 0.2796(2) &  0.337(3)   &  2.1(2) &  6.08(4) &    0.20(2)    &  0.06(3) &  9.6(2)  \\ \hline
		0.9     & 0.2982(2) &  0.336(3)   &  1.9(1) &  5.76(5) &    0.21(2)    &  0.09(1) &  8.1(2)  \\ \hline\hline
	\end{tabular}
\end{indented}
	\caption{\label{tab:results}Asymptotic non-universal and universal quantities obtained for various parameters $\mu$.}
\end{table}

Given the curved geometry in the generated interfaces, we expect the growth to follow the circular KPZ subclass. As previously discussed, the distinction between flat and circular geometry lies in the height (or radius) distribution, with the former governed by the GOE and the latter by the GUE. To precisely assess the agreement with circular KPZ growth, we analyzed the skewness ($S$) and kurtosis ($K$) of the radius distribution.

\begin{figure}[t]
	\begin{indented}
		\item[]	\includegraphics[width=0.5\linewidth]{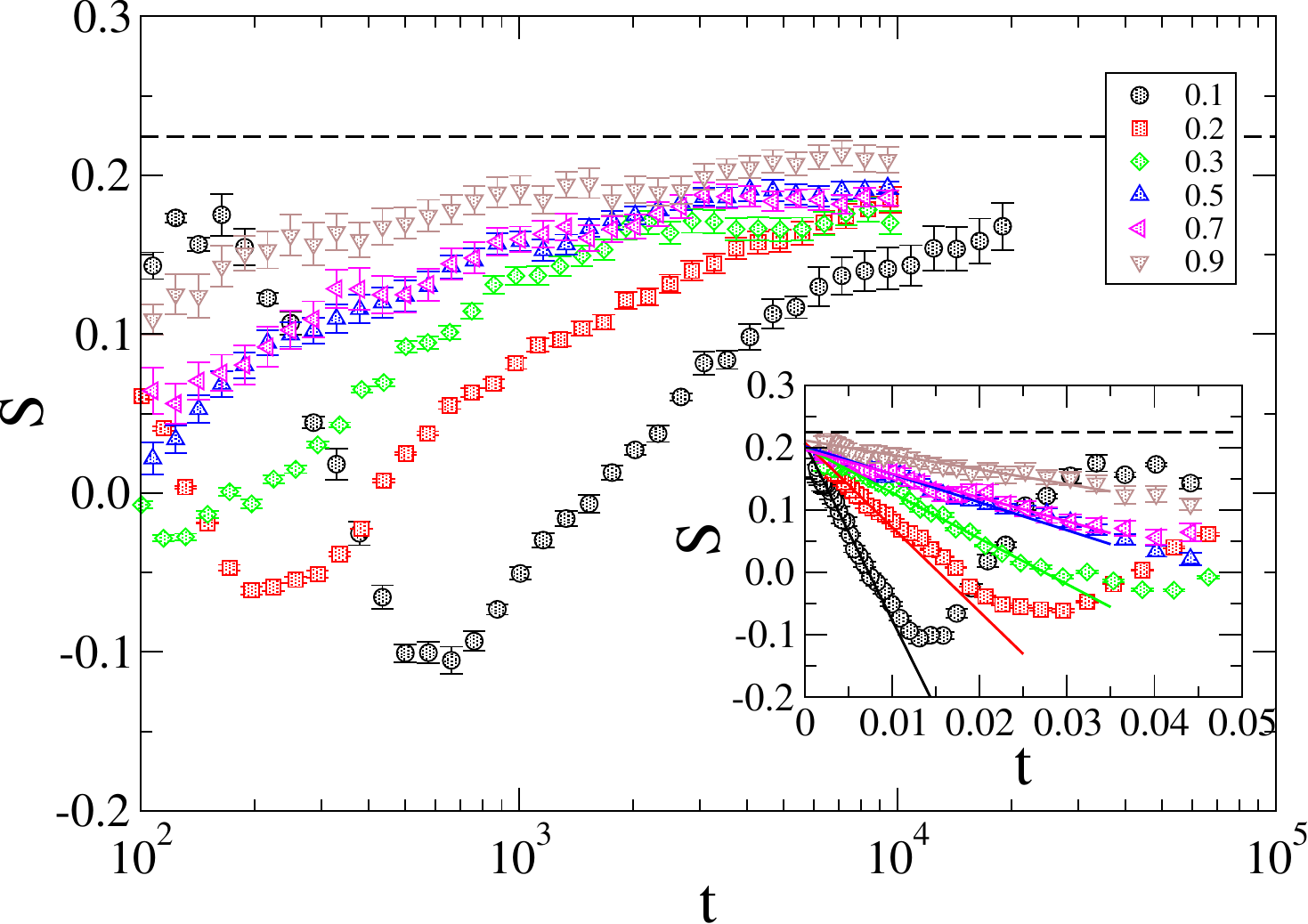} ~~~
		\includegraphics[width=0.49\linewidth]{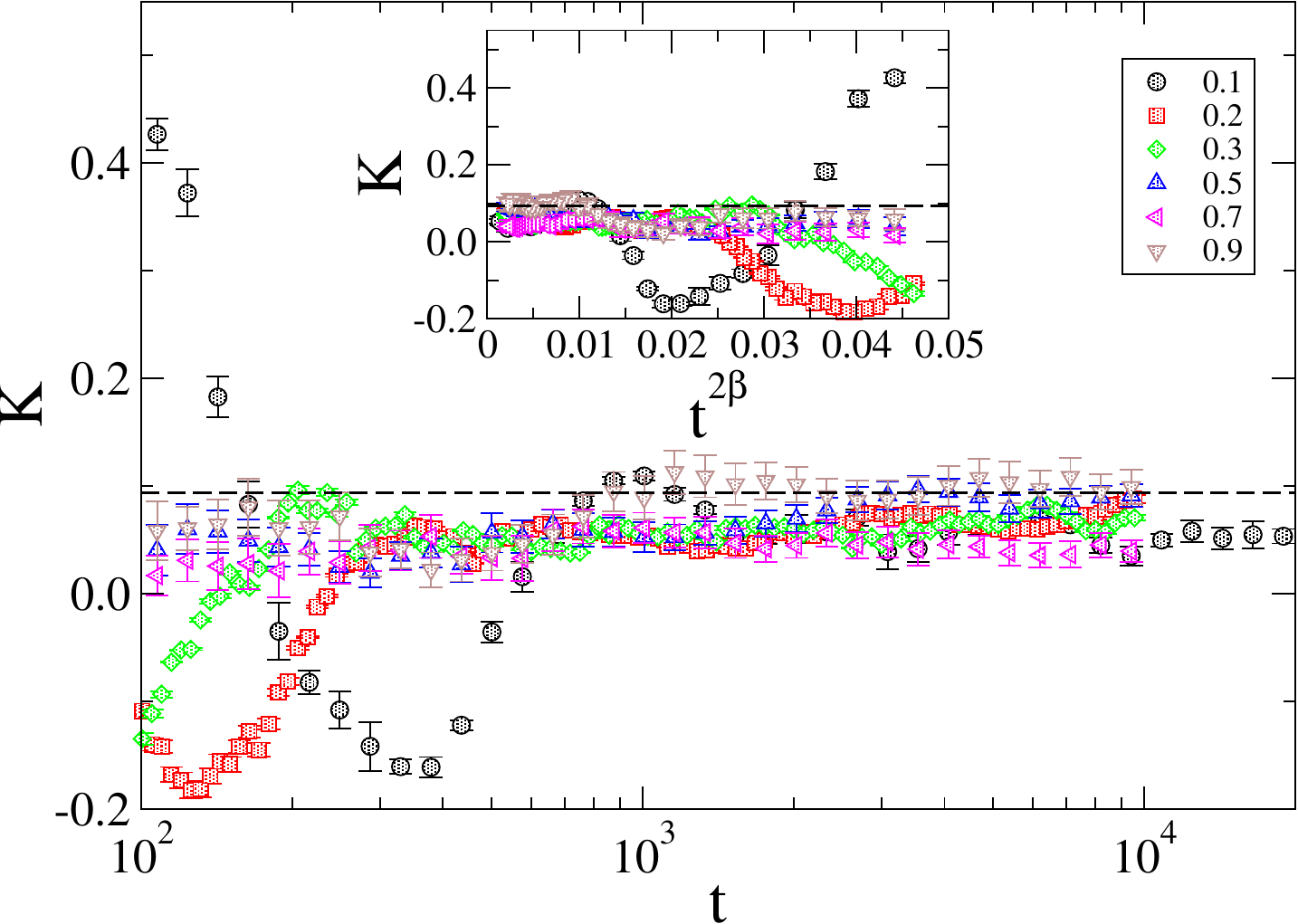}
	\end{indented}
	\caption{\label{fig:SK}Time evolution of skewness and kurtosis for various $\mu$ parameter values (left and right, respectively). The insets show the plots considering $t^{-2\beta}$. The dashed lines are the expected values of skewness and kurtosis for the circular KPZ subclass (S=0.2241 and K=0.09345).} 
\end{figure}

The statistical measures of $S$ and $K$ serve to provide insights into the shape and distribution of the data, offering a means to characterize the behavior and properties of fluctuating variables. The skewness and kurtosis involve the ratio between cumulants and are defined as follows:
\begin{equation}
	S = \frac{\langle{r^3}\rangle_c}{\langle{r^2}\rangle_c^{1.5}} \mbox{~ and ~}
	K = \frac{\langle{r^4}\rangle_c}{\langle{r^2}\rangle_c^2},
\end{equation}
where, we have used $\langle{X^q}\rangle_c$ to represent the cumulant of order $q$.
The main panels in Figure \ref{fig:SK} present the time evolution of the $S$ and $K$. The insets show a plot of $S$ and $K$ against $t^{-a}$, an extrapolation employed to verify the asymptotic skewness and kurtosis values.
The estimated value of $S$ converges linearly to the expected value from the Tracy Widom distribution of the GUE  ($S=0.2241$) for $a=2\beta$ (other values were tested). Differently, in the case of $K$, the values oscillate closely to the expected one ($K=0.09345$). The estimated values are presented in Table \ref{tab:results}. 

\begin{figure}[t]
	\begin{indented}
		\item[]	\includegraphics[width=0.495\linewidth]{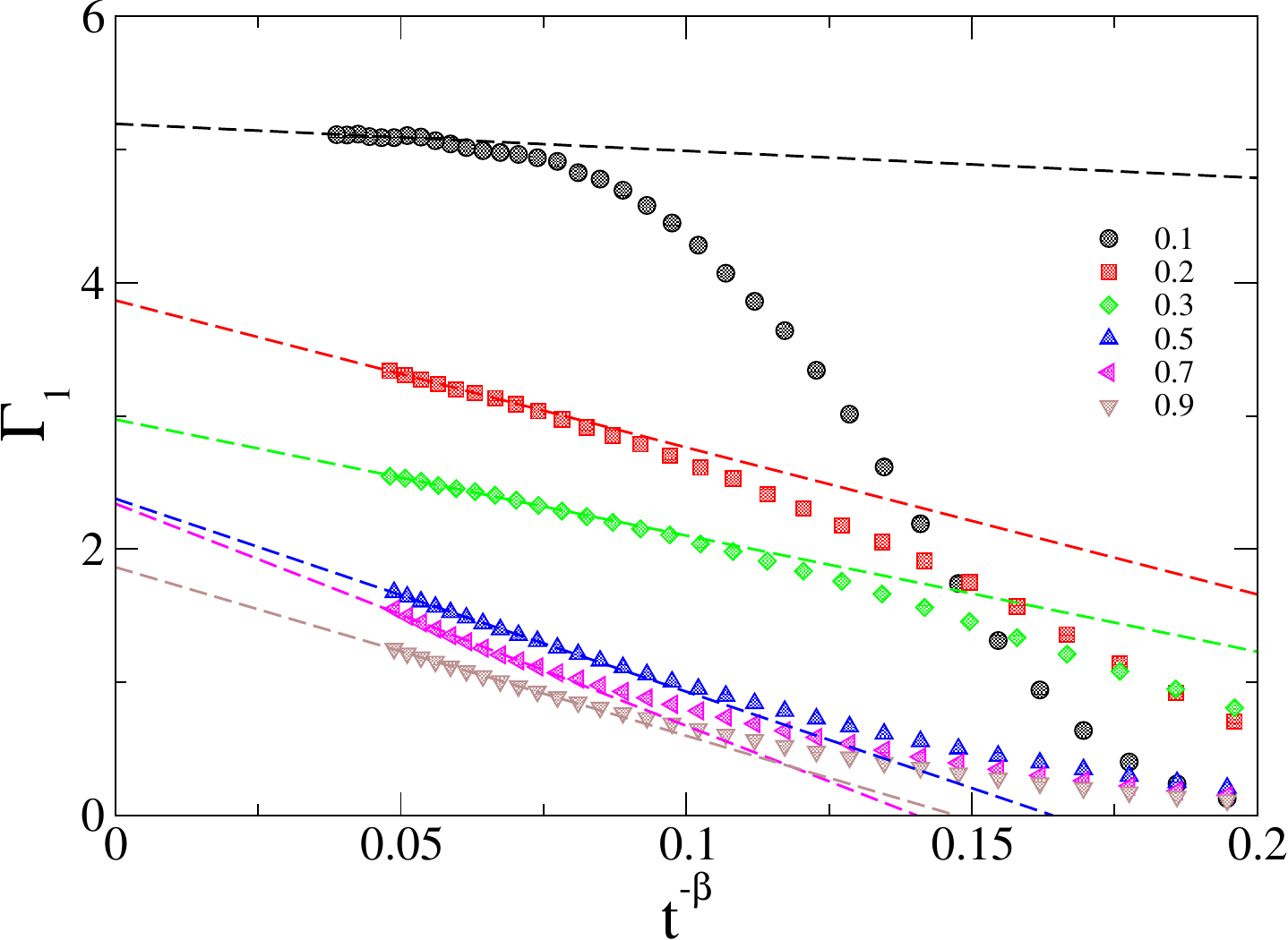}~~~
		\includegraphics[width=0.51\linewidth]{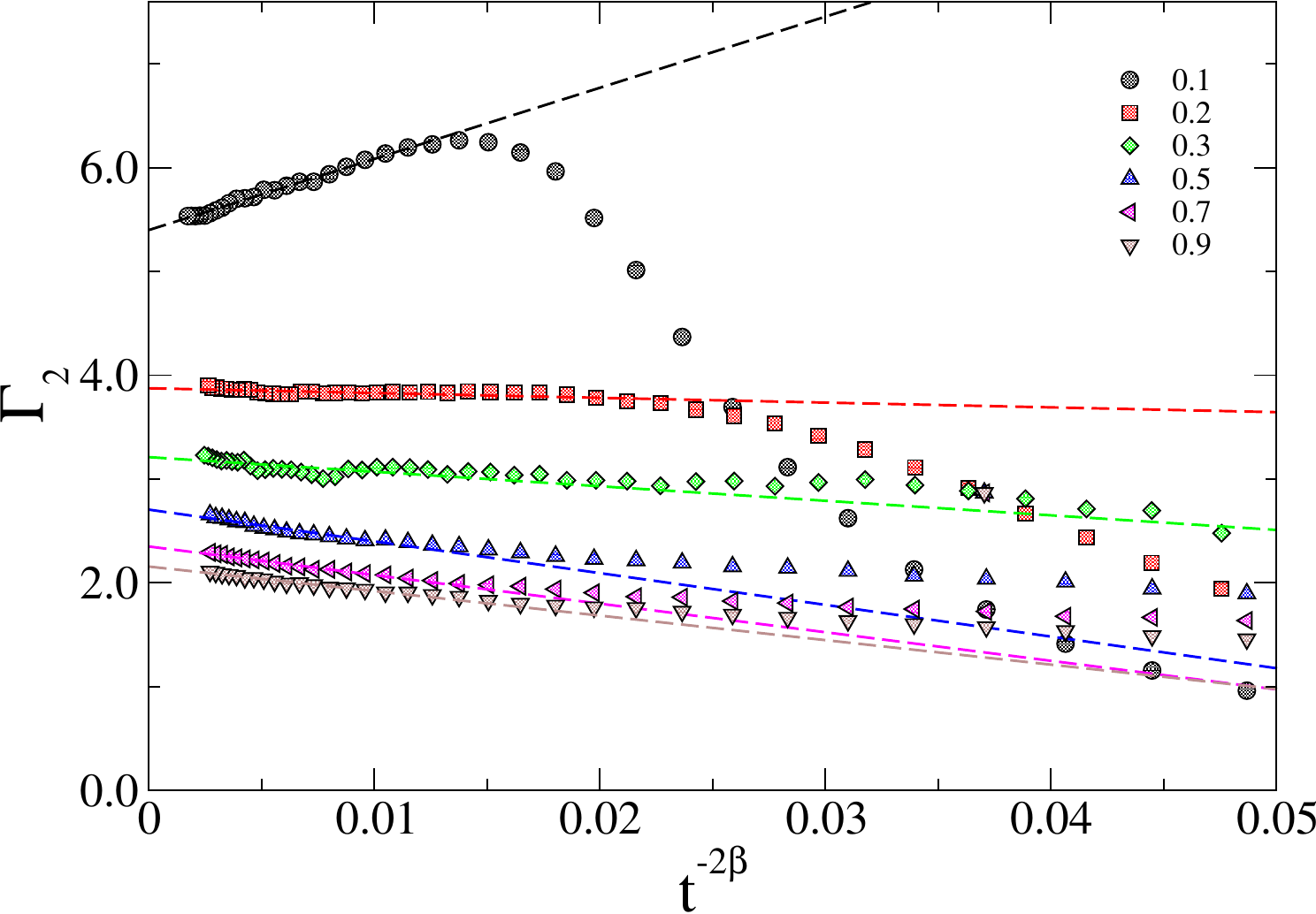}
	\end{indented}
	\caption{\label{fig:Gammas}The non-universal $\Gamma$ parameter estimation. Plot of $\Gamma_1$ (left panel) and $\Gamma_2$ (right panel) against $t^{-\beta}$ and $t^{-2\beta}$, respectively. The dashed lines in both panels represent linear fits used to obtain the extrapolated values.} 
\end{figure}

The next step in our investigation concerns analyzing the properties of the stochastic variable $\chi$. To proceed, we need to determine the parameter $\Gamma$ in Equation \ref{eq:point}. An estimate for $\Gamma$ can be obtained by using the mean radius (Equation \ref{eq:point}) and its second cumulant (Equation \ref{eq:w}) as follows
\begin{equation}
	\Gamma_1^\beta = \frac{\langle r\rangle - v_\infty t}{t^\beta \langle\chi\rangle} \mbox{~ and ~}
	\Gamma_2^{2\beta} = \frac{\langle r^2\rangle_c}{t^{2\beta} \langle\chi^2\rangle_c},
\end{equation}
where $\langle\chi\rangle$ and $\langle\chi^2\rangle_c$ are the first and second cumulants of $\chi$. Since the skewness and the kurtosis values converge to the value of the Tracy Widom distribution of the GUE (Figure \ref{fig:SK}), we adopt the values $\langle\chi\rangle = -1.771069$ and  $\langle\chi^2\rangle_c = 0.812729$. Figure \ref{fig:Gammas} shows the plots of $\Gamma_1$ and $\Gamma_2$ against $t^{-\beta}$ and $t^{-2\beta}$, respectively. We adopt $\Gamma$ as a mean of $\Gamma_1$ and $\Gamma_2$ in the following (see Table \ref{tab:results}).

\begin{figure}[htb]
\begin{indented}
	\item[]	\includegraphics[width=0.65\linewidth]{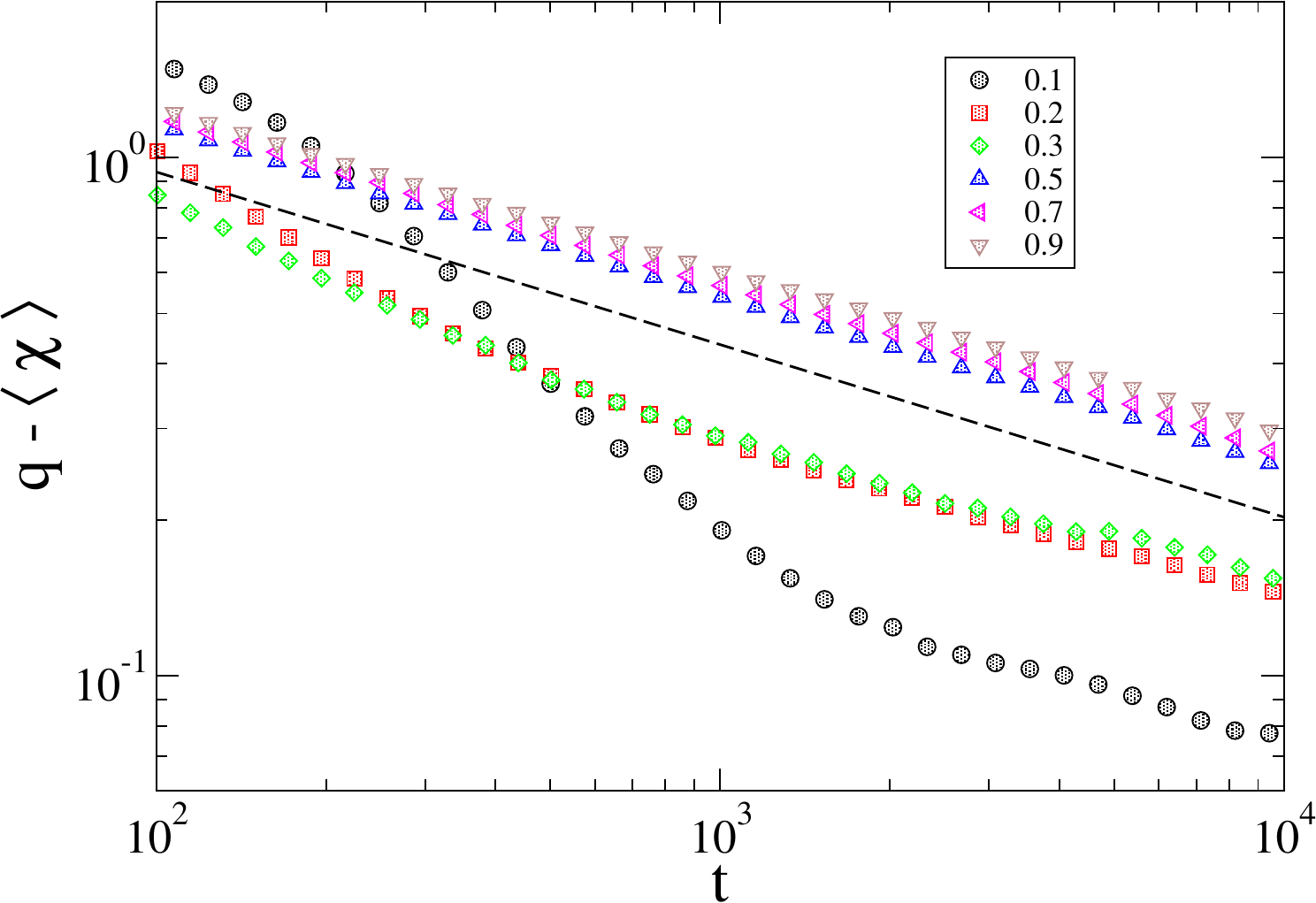}
\end{indented}
	\caption{\label{fig:q_chi} Difference between $q$ and $\chi$ against time. The dashed line is a power law with exponent $-\beta$.}
\end{figure}

Using the previously obtained parameters, we examine the convergence of fluctuations toward the mean value of $\chi$, represented by the random variable $q$. This variable is defined as
\begin{equation}
	q = \frac{\langle h\rangle - v_\infty t }{s_\lambda(\Gamma t)^\beta}.
\end{equation}
In Figure \ref{fig:q_chi}, we plot the difference $q - \langle\chi\rangle$ against time. The plot shows a behavior consistent with a vanishing shift 
with a power law for all values of the $\mu$ parameter. This behavior aligns with the deterministic shift $\eta$ proposed 
in the KPZ ansatz \cite{TakeSano,TakeSano2012,SasaSpohn,Frings,Alves11,Oliveira12}, described by
\begin{equation}
	r_i(t) = v_\infty t + s_\lambda (\Gamma t)^\beta \chi + \eta. \label{eq:point_gen}
\end{equation}
The mean values of the shifts are presented in Table \ref{tab:results}.

\begin{figure}[h]
\begin{indented}
	\item[]	\includegraphics[width=0.5\linewidth]{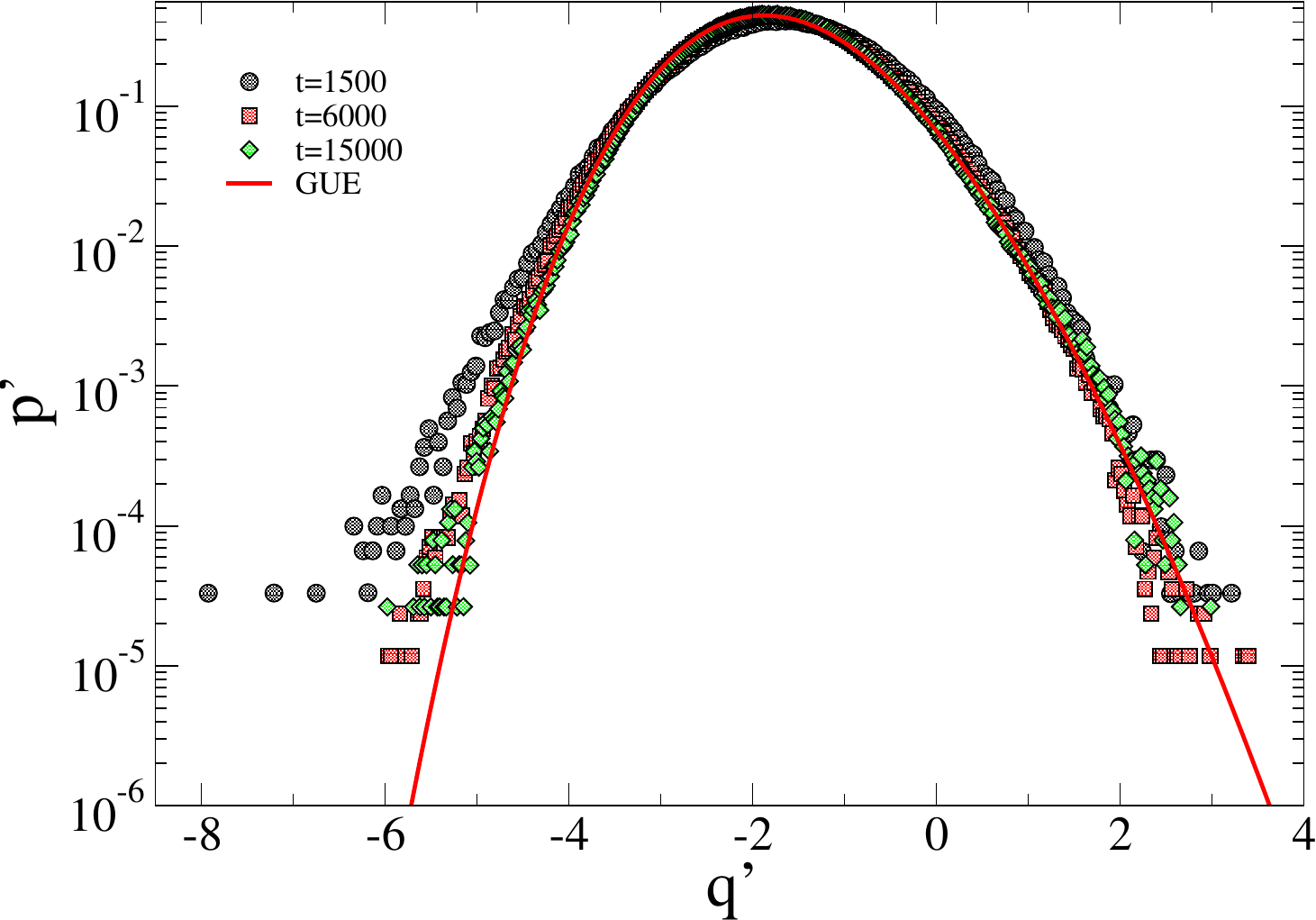} 
	\includegraphics[width=0.5\linewidth]{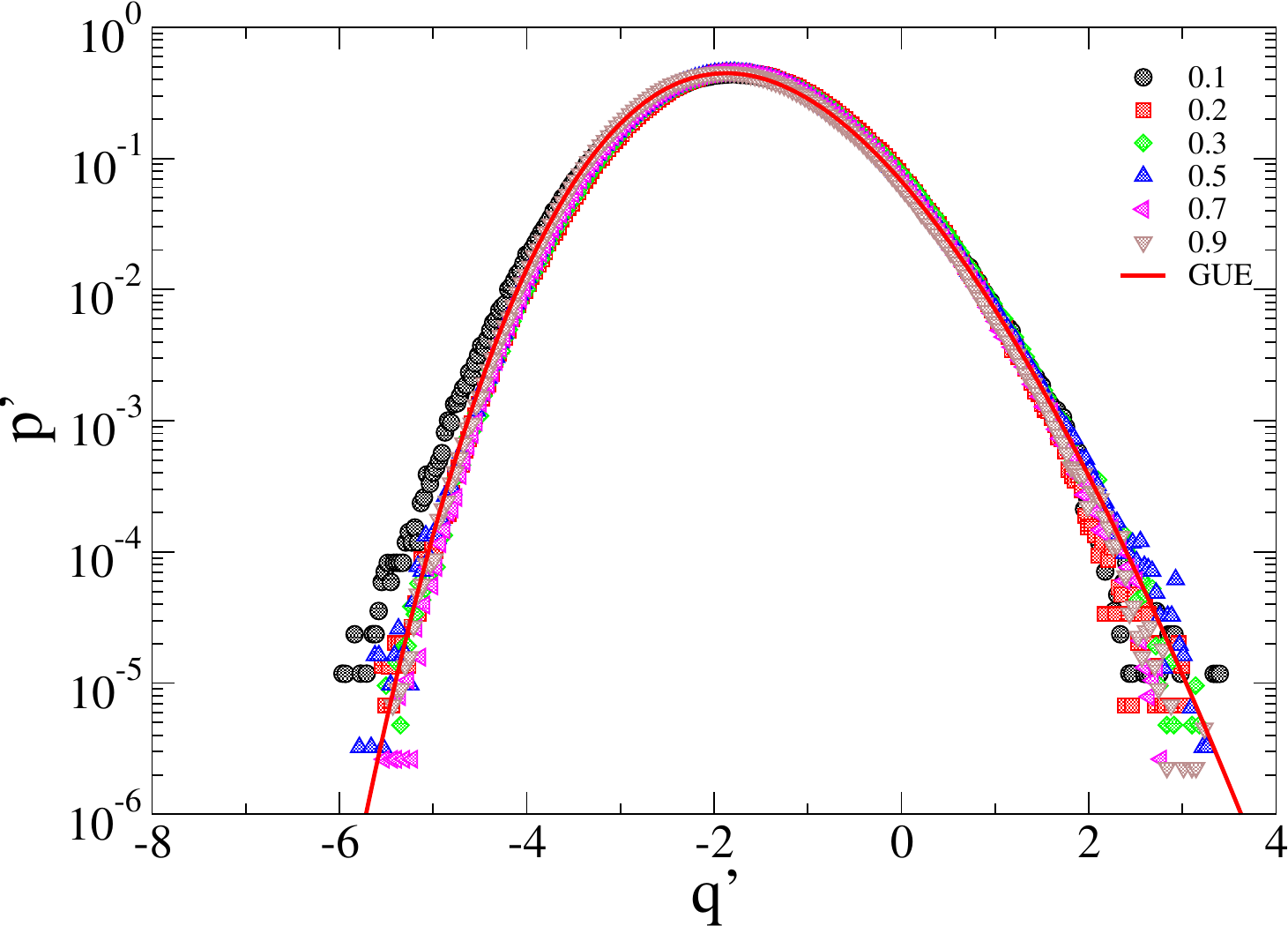}
\end{indented}
	\caption{\label{fig:pq} Distribution probability function for $q'$ (left panel) considering $\mu=0.1$ for three distinct times (right panel) and for different values of $\mu$ at time $1.5 \times 10^4$. The solid lines in both panel are the GUE distribution.}
\end{figure}

Considering the result of the Equation \ref{eq:point_gen}, we can compare the obtained radius distribution function of the particles radii $\{r_i(t)\}$ with that of the Tracy-Widom PDF of GUE expected for the circular KPZ subclass. Therefore, we analyze the distribution function of the rescaled variable $q'$ defined as
\begin{equation}
	q' = \frac{\langle h\rangle - v_\infty t -\eta}{s_\lambda(\Gamma t)^\beta}.
\end{equation}
We used the parameters obtained from Table \ref{tab:results} for this analysis. The left panel of Figure \ref{fig:pq} presents the results for $\mu=0.1$, showing the PDF of the rescaled variable at three different times. The distribution exhibits convergence to the Tracy-Widom PDF of GUE. 
Considering the plots for other values of $\mu$ in the right panel in Figure \ref{fig:pq}, we also observe agreement with the Tracy-Widom PDF of GUE.

\begin{figure}
\begin{indented}
\item[]	\includegraphics[width=0.5\linewidth]{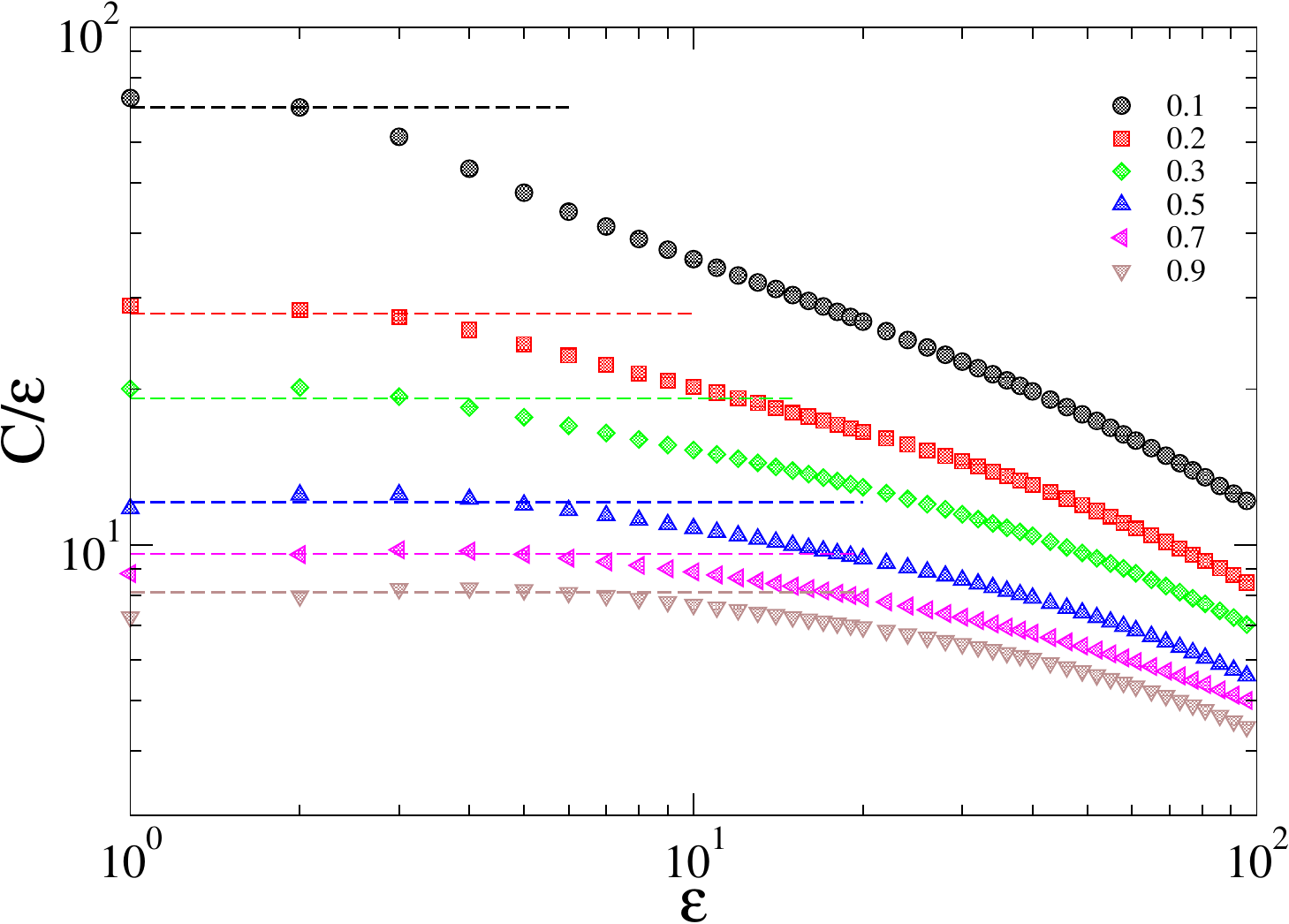}
\includegraphics[width=0.51\linewidth]{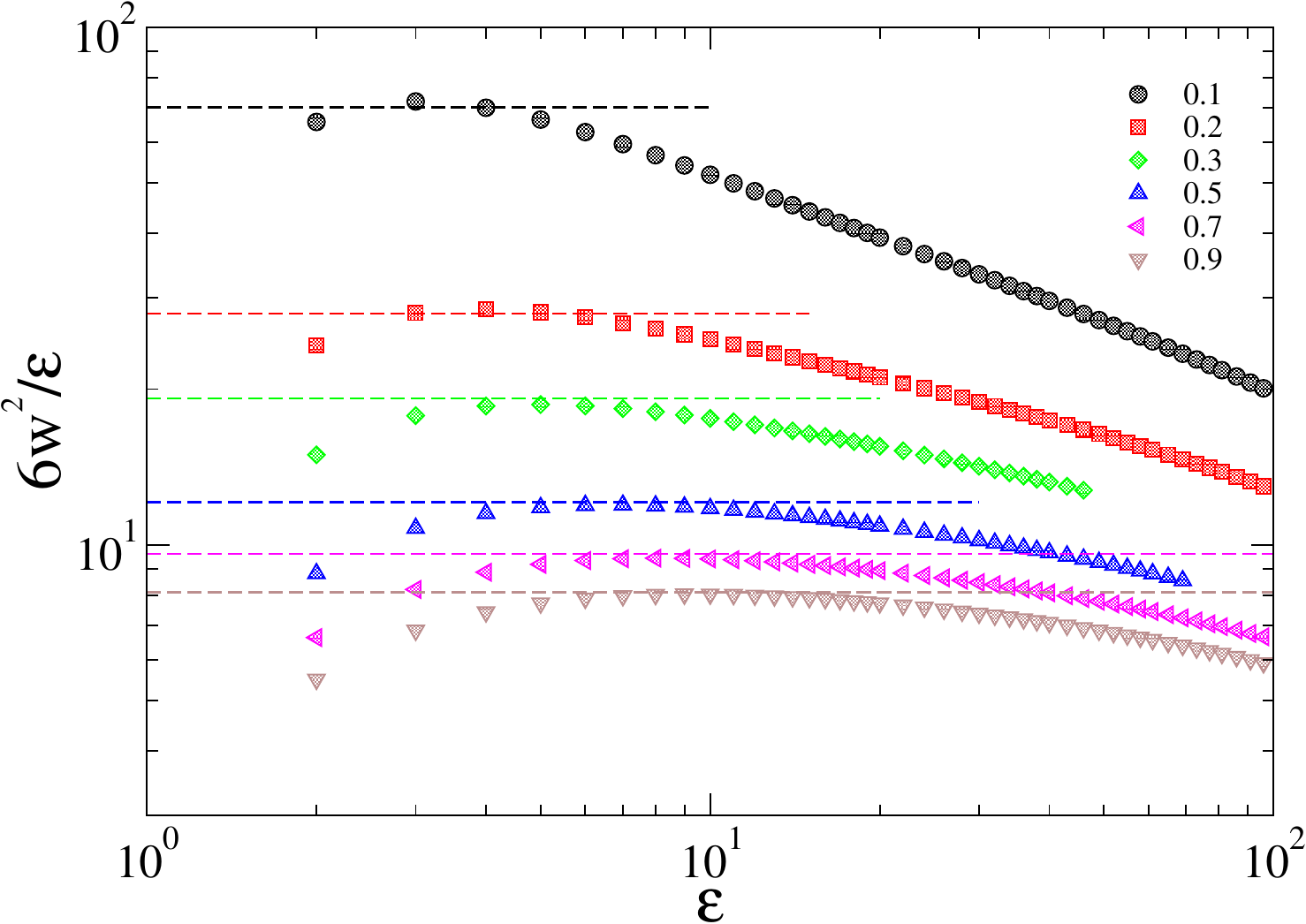}
\end{indented}
\caption{\label{fig:cwe_reesdaled} Estimation of the amplitude $A$ for the various $\mu$ values using the plots of $C(\varepsilon,t)/\varepsilon$ and $w_\varepsilon^2/\varepsilon$ against $\varepsilon$ in left and right, respectively. The dashed lines represent the mean value of the amplitude.}
\end{figure}

\begin{figure}
\begin{indented}
	\item[]\includegraphics[width=0.5\linewidth]{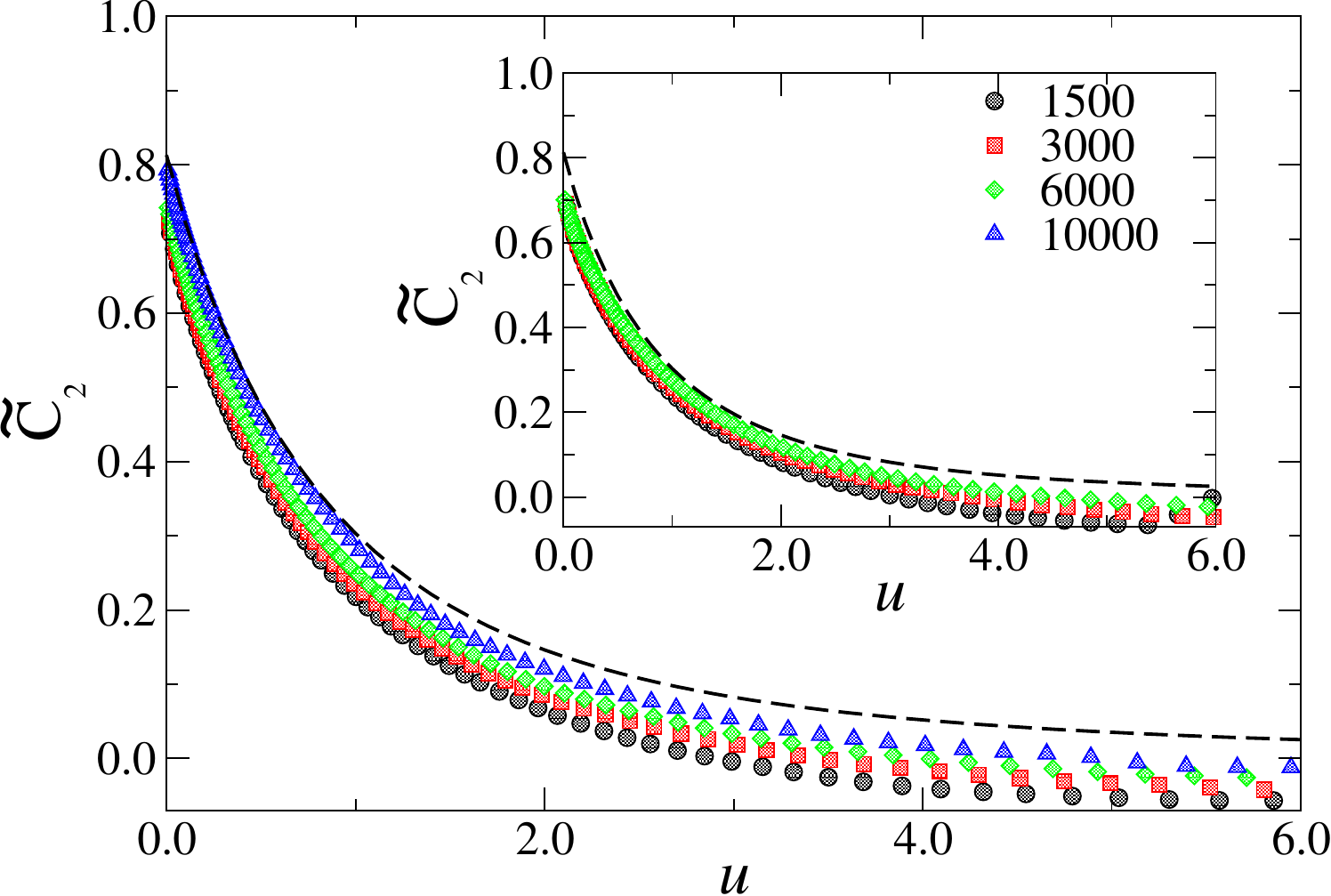}~~~
	\includegraphics[width=0.5\linewidth]{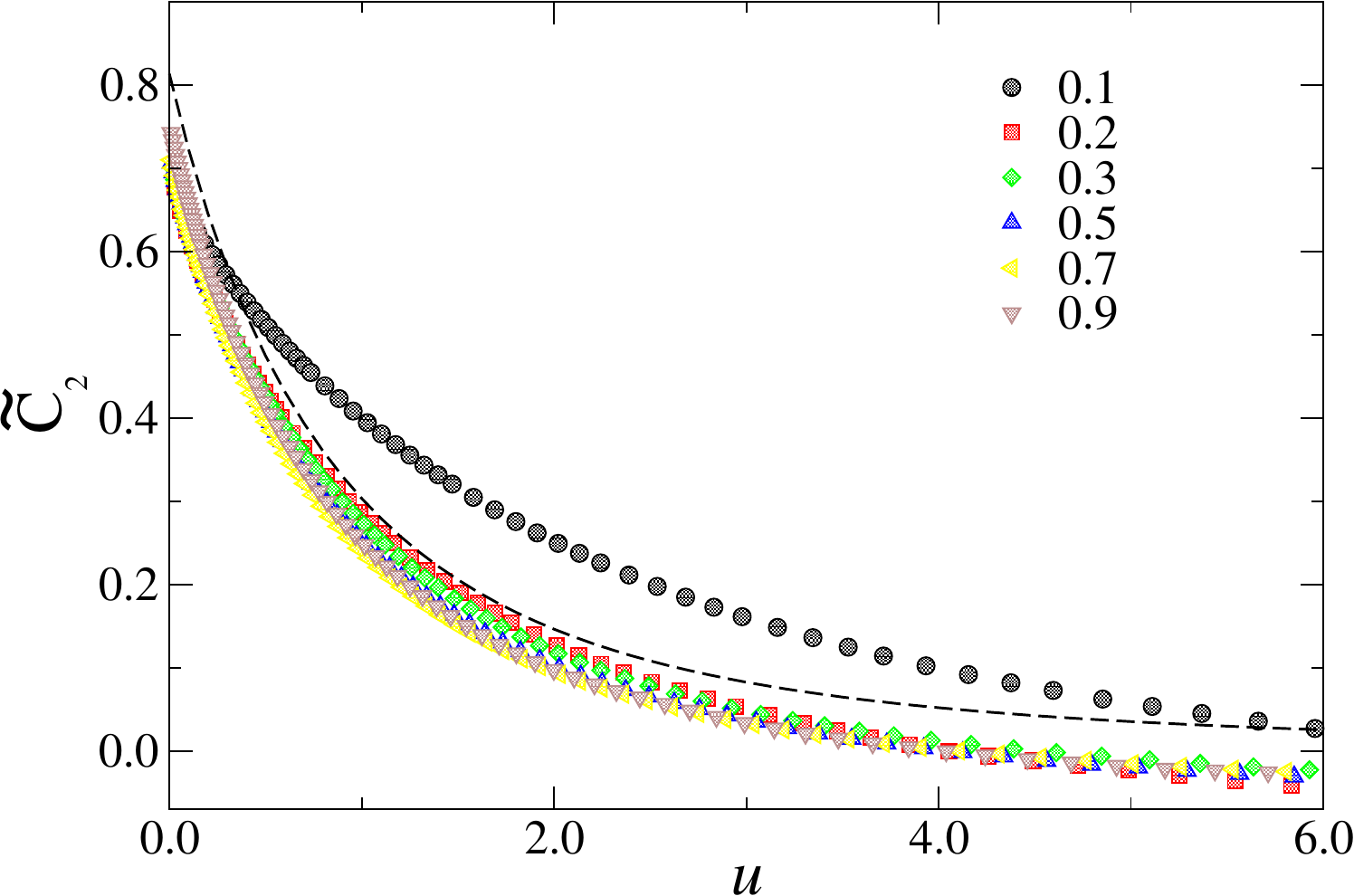}
\end{indented}
\caption{\label{fig:ctil} Scaled two-point correlation function against $u \equiv (A\varepsilon/2)/(\Gamma t)^{2/3}$. Data for four different time points were obtained using $\mu=0.9$ (in the left panel) and different $\mu$ values at time $t=6000$ (right panel). The inset shows the results for $\mu=0.3$ considering $t=1500, 3000,$ and $6000$.}
\end{figure}

Finally, we investigate the two-point correlation function given by
\begin{equation}
	C_2(\varepsilon,t) = \left\langle r_{i+\varepsilon}(t)r_i(t)\right\rangle - \left\langle r\right\rangle.
\end{equation}
For systems belonging to the circular subclass of the KPZ universality, the rescaled two-point correlation function, denoted as $\tilde C_2$ and defined as $C_2/(\Gamma t)^{2\beta}$, against $u = (A\varepsilon/2)(\Gamma ^t)^{2\beta}$ exhibit agreement with the covariance of the Airy$_2$ process\cite{bornemann}. 
To perform this analysis we need to estimate the amplitude $A$. This value can be obtained from the amplitude of the height-difference correlation function $C(\varepsilon,t)$ and the local roughness $w_\varepsilon$ using $C(\varepsilon,t)\simeq A\varepsilon$ and $w_\varepsilon^2\simeq 6 A\varepsilon$ ~\cite{TakeSano,Takeuchi2018}. The analysis is made using the plots of $C(\varepsilon,t)/\varepsilon$ and $w_\varepsilon^2/\varepsilon$ against $\varepsilon$ as shown in Figure \ref{fig:cwe_reesdaled}. We have used the mean value displayed in Table \ref{tab:results}.
We now analyze the rescaling of the two-point correlation function described above. First, we focus on the cases for $\mu=0.3$ and $0.9$.
The results for four different times with $\mu=0.9$ (and three times for $\mu=0.3$) are shown in the left panel (inset of the left panel) of Figure \ref{fig:ctil}. The plots present a convergence to the covariance of the Airy$_2$ process. Except for the case $\mu=0.1$, the other investigated $\mu$ values exhibit a similar behavior (result not shown).
It is important to note that achieving agreement requires significantly longer timescales mainly for the smaller $\mu$ values.
The right panel of Figure \ref{fig:ctil} shows the data for all investigated values of the $\mu$ parameter. As mentioned, the convergence to Airy$_2$ process is worst for smaller values of $\mu$ and in the case $\mu=0.1$ that convergence can not be observed at the reached times in our simulations.

\section{Conclusions}
\label{conclusions}

In the present work, we investigated an off-lattice variant of the $A + A \leftrightarrow A$ reaction-diffusion model with a single seed as the initial condition in a two-dimensional space. Using careful numerical simulations, we investigate the front evolution using the framework from statistical physics. In particular, we consider if the power law of the radius growth, the radius distribution function, and its two-point correlation 
would satisfy the KPZ ansatz. 
The time evolution of one radius point follows the KPZ ansatz and the roughness scales with a growth exponent very close to the 1d KPZ one for all values of the parameter $\mu$ (that control the division and movement). The cumulants and their ratios obtained for the radius distribution of the front agree among the different values of $\mu$  with those expected for the circular subclass of the KPZ universality. The results for the two-point correlation function exhibit a very slow convergence. The main results indicate that agreement requires significantly longer timescales.

\section*{Acknowledgments}
We thank Ismael S. S. Carrasco for the discussions and critical manuscript reading.
This work was partially supported by CNPq and FAPEMIG (Brazilian agencies).

\section*{References}
\bibliographystyle{iopart-num}
\bibliography{reaction_diffusion_off}

\providecommand{\newblock}{}
\begin{thebibliography}{10}
\expandafter\ifx\csname url\endcsname\relax
  \def\url#1{{\tt #1}}\fi
\expandafter\ifx\csname urlprefix\endcsname\relax\def\urlprefix{URL }\fi
\providecommand{\eprint}[2][]{\url{#2}}

\bibitem{Riordan1995}
Riordan J, Doering C~R and Ben-Avraham D 1995 {\em Physical Review Letters\/}
  {\bf 75} 565--568
  \urlprefix\url{https://link.aps.org/doi/10.1103/PhysRevLett.75.565}

\bibitem{barabasi}
Barabasi A~L and Stanley H~E 1995 {\em Fractal Concepts in Surface Growth\/}
  (Cambridge, England: Cambridge University Press)

\bibitem{meakin}
Meakin P 1998 {\em Fractals, Scaling and Growth far from Equi\-li\-brium\/}
  (Cambridge, England: Cambridge University Press)

\bibitem{BenAvraham1990}
Ben-Avraham D, Burschka M~A and Doering C~R 1990 {\em Journal of Statistical
  Physics\/} {\bf 60} 695--728

\bibitem{Tripathy2000}
Tripathy G and van Saarloos W 2000 {\em Physical Review Letters\/} {\bf 85}
  3556--3559
  \urlprefix\url{https://link.aps.org/doi/10.1103/PhysRevLett.85.3556}

\bibitem{Tripathy2001}
Tripathy G, Rocco A, Casademunt J and van Saarloos W 2001 {\em Physical Review
  Letters\/} {\bf 86} 5215--5218
  \urlprefix\url{https://link.aps.org/doi/10.1103/PhysRevLett.86.5215}

\bibitem{Moro2001}
Moro E 2001 {\em Physical Review Letters\/} {\bf 87} 238303
  \urlprefix\url{https://link.aps.org/doi/10.1103/PhysRevLett.87.238303}

\bibitem{Moro2004}
Moro E 2004 {\em Physical Review E\/} {\bf 69} 060101
  \urlprefix\url{https://link.aps.org/doi/10.1103/PhysRevE.69.060101}

\bibitem{Nesic2014}
Nesic S, Cuerno R and Moro E 2014 {\em Physical Review Letters\/} {\bf 113}
  180602
  \urlprefix\url{https://link.aps.org/doi/10.1103/PhysRevLett.113.180602}

\bibitem{KPZ}
Kardar M, Parisi G and Zhang Y~C 1986 {\em Phys. Rev. Lett.\/} {\bf 56}
  889--892 \urlprefix\url{https://link.aps.org/doi/10.1103/PhysRevLett.56.889}

\bibitem{Almeida2014}
Almeida R~A~L, Ferreira S~O, Oliveira T~J and Reis F~D~A~A~a 2014 {\em Phys.
  Rev. B\/} {\bf 89}(4) 045309
  \urlprefix\url{https://link.aps.org/doi/10.1103/PhysRevB.89.045309}

\bibitem{Almeida2015}
Almeida R~A~L, Ferreira S~O, Ribeiro I~R~B and Oliveira T~J 2015 {\em
  Europhysics Letters\/} {\bf 109} 46003
  \urlprefix\url{https://dx.doi.org/10.1209/0295-5075/109/46003}

\bibitem{Huergo2010}
Huergo M~A~C, Pasquale M~A, Bolz\'an A~E, Arvia A~J and Gonz\'alez P~H 2010
  {\em Phys. Rev. E\/} {\bf 82}(3) 031903
  \urlprefix\url{https://link.aps.org/doi/10.1103/PhysRevE.82.031903}

\bibitem{Huergo2011}
Huergo M~A~C, Pasquale M~A, Gonz\'alez P~H, Bolz\'an A~E and Arvia A~J 2011
  {\em Phys. Rev. E\/} {\bf 84}(2) 021917
  \urlprefix\url{https://link.aps.org/doi/10.1103/PhysRevE.84.021917}

\bibitem{Huergo2012}
Huergo M~A~C, Pasquale M~A, Gonz\'alez P~H, Bolz\'an A~E and Arvia A~J 2012
  {\em Phys. Rev. E\/} {\bf 85}(1) 011918
  \urlprefix\url{https://link.aps.org/doi/10.1103/PhysRevE.85.011918}

\bibitem{johansson}
Johansson K 2000 {\em Commun. Math. Phys\/} {\bf 209} 437--476

\bibitem{PraSpo1}
Pr\"ahofer M and Spohn H 2000 {\em Phys. Rev. Lett.\/} {\bf 84} 4882--4885
  \urlprefix\url{https://link.aps.org/doi/10.1103/PhysRevLett.84.4882}

\bibitem{krug92}
Krug J, Meakin P and Halpin-Healy T 1992 {\em Phys. Rev. A\/} {\bf 45}(2)
  638--653 \urlprefix\url{https://link.aps.org/doi/10.1103/PhysRevA.45.638}

\bibitem{PraSpo2}
Pr\"ahofer M and Spohn H 2000 {\em Physica A\/} {\bf 279} 342--352

\bibitem{Corwin2012}
Corwin I 2012 {\em Random Matrices: Theory and Applications\/} {\bf 1} 1130001
  \urlprefix\url{DOI:10.1142/S2010326311300014}

\bibitem{Takeuchi2018}
Takeuchi K~A 2018 {\em Physica A: Statistical Mechanics and its Applications\/}
  {\bf 504} 77--105 \urlprefix\url{https://doi.org/10.1016/j.physa.2018.03.009}

\bibitem{SasaSpohn}
Sasamoto T and Spohn H 2010 {\em Phys. Rev. Lett.\/} {\bf 104}(23) 230602
  \urlprefix\url{https://link.aps.org/doi/10.1103/PhysRevLett.104.230602}

\bibitem{Amir}
Amir G, Corwin I and Quastel J 2011 {\em Commun. Pure Appl. Math.\/} {\bf 64}
  466--537

\bibitem{Calabrese}
Calabrese P and Le~Doussal P 2011 {\em Phys. Rev. Lett.\/} {\bf 106}(25) 250603
  \urlprefix\url{https://link.aps.org/doi/10.1103/PhysRevLett.106.250603}

\bibitem{Imamura}
Imamura T and Sasamoto T 2012 {\em Phys. Rev. Lett.\/} {\bf 108}(19) 190603
  \urlprefix\url{https://link.aps.org/doi/10.1103/PhysRevLett.108.190603}

\bibitem{TakeSano}
Takeuchi K~A and Sano M 2010 {\em Phys. Rev. Lett.\/} {\bf 104} 230601
  \urlprefix\url{https://link.aps.org/doi/10.1103/PhysRevLett.104.230601}

\bibitem{TakeuchiSP}
Takeuchi K~A, Sano M, Sasamoto T and Spohn H 2011 {\em Sci. Rep.\/} {\bf 1} 34

\bibitem{TakeSano2012}
Takeuchi K~A and Sano M 2012 {\em Journal of Statistical Physics\/} {\bf 147}
  853--890

\bibitem{Alves11}
Alves S~G, Oliveira T~J and Ferreira S~C 2011 {\em Europhys. Lett.\/} {\bf 96}
  48003

\bibitem{Oliveira12}
Oliveira T~J, Ferreira S~C and Alves S~G 2012 {\em Phys. Rev. E\/} {\bf 85}
  010601 \urlprefix\url{https://link.aps.org/doi/10.1103/PhysRevE.85.010601}

\bibitem{Alves13}
Alves S~G, Oliveira T~J and Ferreira S~C 2013 {\em Journal of Statistical
  Mechanics: Theory and Experiment\/} {\bf 2013} P05007

\bibitem{Carrasco_2014}
Carrasco I~S~S, Takeuchi K~A, Ferreira S~C and Oliveira T~J 2014 {\em New
  Journal of Physics\/} {\bf 16} 123057
  \urlprefix\url{https://dx.doi.org/10.1088/1367-2630/16/12/123057}

\bibitem{Healy14}
Halpin-Healy T and Lin Y 2014 {\em Phys. Rev. E\/} {\bf 89}(1) 010103
  \urlprefix\url{https://link.aps.org/doi/10.1103/PhysRevE.89.010103}

\bibitem{Santalla_2015}
Santalla S~N, guez Laguna J~R, LaGatta T and Cuerno R 2015 {\em New Journal of
  Physics\/} {\bf 17} 033018
  \urlprefix\url{https://dx.doi.org/10.1088/1367-2630/17/3/033018}

\bibitem{HealyTake}
Halpin-Healy T and Takeuchi K~A 2015 {\em Journal of Statistical Physics\/}
  {\bf 160} 638

\bibitem{Santalla_2017}
Santalla S~N, Rodríguez-Laguna J, Celi A and Cuerno R 2017 {\em Journal of
  Statistical Mechanics: Theory and Experiment\/} {\bf 2017} 023201
  \urlprefix\url{https://dx.doi.org/10.1088/1742-5468/aa5754}

\bibitem{Alves2018}
Alves S 2018 {\em Phys. Rev. E\/} {\bf 97} 032801
  \urlprefix\url{https://link.aps.org/doi/10.1103/PhysRevE.97.032801}

\bibitem{Roy}
Roy D and Pandit R 2020 {\em Phys. Rev. E\/} {\bf 101}(3) 030103
  \urlprefix\url{https://link.aps.org/doi/10.1103/PhysRevE.101.030103}

\bibitem{Barreales_2020}
Barreales B~G, Mel{\'{e}}ndez J~J, Cuerno R and Ruiz-Lorenzo J~J 2020 {\em
  Journal of Statistical Mechanics: Theory and Experiment\/} {\bf 2020} 023203
  \urlprefix\url{https://doi.org/10.1088/1742-5468/ab6a03}

\bibitem{AlvesBJP}
Alves S~G, Ferreira S~C and Martins M~L 2008 {\em Braz. J. Phys.\/} {\bf 38}
  81--86

\bibitem{Frings}
Ferrari P and Frings R 2011 {\em J. Stat. Phys.\/} {\bf 144} 1--28

\bibitem{bornemann}
 2008 {\em Journal of Statistical Physics\/} {\bf 133} 405--415 ISSN 0022-4715
  \urlprefix\url{http://link.springer.com/10.1007/s10955-008-9621-0}

\end{thebibliography}

\end{document}